\documentclass[twocolumn,showpacs,floatfix,superscriptaddress]{revtex4-1}

\usepackage{amssymb,amsmath,amstext}                
\usepackage{graphicx}                                               
\usepackage{epstopdf}                                               
\usepackage{color}       
\usepackage{bm}
\usepackage{appendix}                                              
\usepackage[utf8]{inputenc}
\usepackage{bbold}
\usepackage{bbm}
\usepackage{latexsym}
\usepackage{mathrsfs}
\usepackage[colorlinks=true,citecolor=blue,linkcolor=magenta]{hyperref}

\begin{document}
\title{Finite correlation length scaling with infinite projected entangled pair states at finite temperature}

\author{Piotr Czarnik}
\affiliation{Institute of Nuclear Physics, Polish Academy of Sciences, Radzikowskiego 152, PL-31342 Krak\'ow, Poland}

\author{ Philippe Corboz }
\affiliation{Institute for Theoretical Physics and Delta Institute for Theoretical Physics, 
             University of Amsterdam, Science Park 904, 1098 XH Amsterdam, The Netherlands}

\date{\today}

\begin{abstract}
We study  second order finite temperature phase transitions of the 2D  quantum Ising and interacting honeycomb fermions models using infinite projected entangled pair states (iPEPS). We obtain  an iPEPS thermal state representation  by Variational Tensor Network Renormalization (VTNR). We find that  at the critical temperature $T_c$  the iPEPS correlation length is finite for the computationally accessible values of the iPEPS bond dimension $D$. Motivated by this observation we  investigate the application of   Finite Correlation Length Scaling (FCLS), which has been previously used for iPEPS  simulations of quantum critical points at $T=0$,  to obtain precise values of  $T_c$  and the universal critical exponents. We find that in the vicinity of $T_c$ the behavior of  observables follows well the one predicted by FCLS. Using FCLS we obtain  $T_c$ and the  critical exponents in agreement  with Quantum Monte Carlo (QMC) results except for couplings close to the quantum critical points where larger bond dimensions are required.

\end{abstract}
\maketitle


\section{Introduction}

Tensor networks \cite{Verstraete_review_08, Schollwock_review_11, Orus_review_14, Bridgeman_rev_17,Orus_rev_18} are representations of weakly  entangled  states obeying an  area law of entanglement \cite{Hastings_GSarealaw_07,Eisert_rev_10,Laflorencie_rev_16}. They are a basis for variational numerical methods  for strongly correlated quantum many body systems, enabling simulations of fermionic, bosonic and spin models with the same leading computational complexity \cite{Eisert_fMERA_09,Corboz_fMERA_10,Corboz_fMERA_09,Cirac_fPEPS_10,Barthel_fTN_09,Corboz_fiPEPS_10}. The powerful density matrix renormalization group (DMRG) \cite{White_DMRG_92, White_DMRG_93} approximates a state of a system by  a 1D tensor network called  matrix product state (MPS) \cite{AKLT,Fannes_MPS_92,Ostlund_MPS_95}. 2D projected entangled pair  states (PEPS) \cite{Verstraete_PEPS_04}, called  also tensor product states \cite{Nishino_2DvarTN_01,Nishino_2DvarTN_04}, were initially applied  as a variational ansatz for 2D ground states \cite{Murg_finitePEPS_07,Cirac_iPEPS_08,Xiang_SU_08,Corboz_fiPEPS_10,Cirac_fPEPS_10}  bringing new insights into paradigmatic models of strongly correlated systems (see e. g. \cite{Corboz_CTM_14,Simons_Hubb_17,Xiang_pess_14,Xiang_kagome_17,Poilblanc_iPEPScrit_17,Haghshenas_u1_18,Corboz_SS_14}). Recent years brought  new applications of PEPS \cite{Kshetrimayum_diss_17,Vanderstraeten_PEPSext_18,Kennes_MBLPEPS_18,Czarnik_FU_19,Hubig_PEPSev_18} and further progress in the fields of numerical optimization \cite{Lubasch_gauge_14,Phien_ffu_15,Corboz_varopt_16,Vanderstraeten_varopt_16} and contraction \cite{Xie_PEPScontr_17,Fishman_FPCTM_17} of PEPS.

Thermal states of 2D  local Hamiltonians obey an area law for mutual information, which is reproduced by projected entangled-pair  operators (iPEPO) representing thermal states  and  infinite projected entangled-pair  states (iPEPS) representing purifications  of thermal states  ~\cite{Wolf_Tarealaw_08}, giving motivation to use iPEPS  for thermal states simulations \cite{Czarnik_evproj_12,Czarnik_fevproj_14}. Recently new methods for  simulation of  thermal states, based on   iPEPS  and iPEPO, were proposed 
\cite{Czarnik_VTNR_15, Czarnik_fVTNR_16,  Czarnik_var_18, Orus_SUfiniteT_18, Czarnik_FU_19}. Recent years brought also developments in the field of the closely related direct contraction methods for 3D tensor networks  representing partition functions  of 2D quantum models \cite{Gu_TERG_08,Li_LTRG_11, Xie_HOSRG_12, Ran_ODTNS_12,  Ran_NCD_13} and MPS/MPO based  simulations of  thermal states of finite width cylinders \cite{Stoudenmire_2DMETTS_17, Weichselbaum_Tdec_18}. Some of those methods were already applied to challenging problems \cite{Su_THAFoctakagome_17, Czarnik_eg_17, Ran_THAFstar_18,  Su_THAFkagome_17, Weichselbaum_tHAF_18}.

 Among demanding
problems in the field of 2D strongly correlated systems are finite temperature critical phenomena and in particular finite temperature second order phase transitions. Some of the methods mentioned above were already applied to investigate   2D critical phenomena based on the assumption that large enough $D$ can be obtained to  provide results which are converged in~$D$~\cite{Su_THAFoctakagome_17, Czarnik_eg_17,  Su_THAFkagome_17, Czarnik_FU_19}, however, for more challenging cases reaching convergence in $D$ will in general be difficult.

 Here we demonstrate that even in the case when  convergence in $D$  cannot be obtained,  it is possible to take  finite $D$ effects systematically into account using a Finite Correlation Length Scaling (FCLS)~\cite{Tagliacozzo_FES_08,Pollmann_FES_09,Corboz_FCLS_18,Rader_FCLS_18}. Furthermore, we show that FCLS  can be used  to obtain critical data for a finite temperature phase transition, i. e. the critical temperature $T_c$ and the universal critical exponents.

FCLS, originally called  finite entanglement scaling (FES), was first proposed to investigate 1D quantum critical points by infinite MPS (iMPS)~\cite{Tagliacozzo_FES_08,Pollmann_FES_09,Pirvu_FES_12}. These critical points  violate the area law of entanglement~\cite{Vidal_1Dcrit_03} and as such cannot be represented by finite $D$ iMPS, which have a finite correlation length $\xi_D$. It was shown that in the case of  the optimal iMPS finite $D$  ground state approximation the finite $D$ modifies observables of the critical state  as if the system was finite with the size proportional to  $\xi_{D}$~\cite{Tagliacozzo_FES_08,Pollmann_FES_09}.   It was also shown that  the scaling of the observables with  increasing $\xi_{D}$ can be used to  determine the critical exponents and the precise location of the critical point similarly as in standard finite size scaling for Quantum Monte Carlo (QMC)  simulations~\cite{Tagliacozzo_FES_08,Pollmann_FES_09,Pirvu_FES_12}.

A similar idea was  applied earlier in corner transfer matrix renormalization group (CTMRG) simulations of 2D critical thermal states of classical models. CTMRG approximately contracts  a 2D tensor network  representing a  partition function of a 2D classical system. The approximation introduces  an effective correlation length controlled by a refinement parameter $\chi$ of the method~\cite{Nishino_scalCTM_96}. A scaling ansatz assuming that this correlation length is proportional to  the effective system size was introduced to find the critical properties~\cite{Nishino_scalCTM_96}. 

FCLS was recently applied to  iPEPS simulations of  2D Lorentz-invariant quantum critical points,  i. e. quantum critical points with a linear dispersion relation of low energy excitations~\cite{Corboz_FCLS_18,Rader_FCLS_18}. It was shown that in such case  the optimal  finite $D$ iPEPS approximating a critical ground state has a finite correlation length $\xi_D$, and that FCLS can be used  to determine the critical coupling and the universal critical exponents. 

In this paper we simulate second order finite temperature  phase transitions for a  2D quantum Ising model and interacting spinless fermions on a honeycomb lattice using Variational Tensor Network Renormlization (VTNR) \cite{Czarnik_VTNR_15, Czarnik_fVTNR_16}. For thermal states at finite $T$ we can expect that the exact state can be represented with a finite bond dimension $D_{exact}$~\cite{Corboz_FCLS_18}. Here we find that at the critical temperature the obtained thermal states have a finite correlation length $\xi_D$  for all  bond dimensions used in this work suggesting that we are in a regime where $D<D_{exact}$. This motivates us to investigate the  possibility to use FCLS also in these cases. In this paper we present benchmark results demonstrating that indeed FCLS can be applied to determine the critical data. 

This paper is organized as follows: In Sec.~\ref{sec:puri} we introduce a thermal state's representation by a purification and in Sec. \ref{sec:iPEPS} we describe how to represent such a purification using an iPEPS. In Sec.~\ref{sec:VTNR}  we introduce VTNR which we use to obtain the  purification's iPEPS representation. In Sec.~\ref{sec:CTM} we describe the CTMRG method which allows us  to efficiently contract a 2D tensor network. In Sec.~\ref{sec:extrap} we explain how to determine the correlation length of an iPEPS using CTMRG. In Sec.~\ref{sec:FCLS} we introduce FCLS and in Sec.~\ref{sec:Tc} we describe how to determine the  critical temperature  $T_c$ for a second order phase transition using  FCLS.  
In Secs.~\ref{sec:QIs} and \ref{sec:honfer} we present the benchmark results for the application of FCLS to simulations of thermal second order phase transitions in the quantum Ising and  interacting honeycomb fermions models, respectively. Finally, we provide our conclusions in Sec.~\ref{sec:conclusions}.    

\section{Purifications  of thermal states} 
\label{sec:puri}

A thermal state of a Hamiltonian $H$ for temperature $T$ is given by its thermal density matrix
\begin{equation}
\rho(T) = \frac{1}{Z(T)} e^{-H/T}, \quad Z(T) = \textrm{Tr}\, e^{-H/T}.
\end{equation}
Here we consider lattice models for which the Hilbert space $\mathcal{H}$  is a tensor product  of Hilbert spaces of individual lattice sites $\mathcal{H}_i$ spanned by states $\{ |s_i\rangle, s_i = 1\dots d \}$.

 We represent $\rho(T)$ by its purification $|\Psi(T)\rangle$ which is a pure state in an enlarged Hilbert space $\tilde{\mathcal{H}}$  created by introducing  ancillary degrees of freedom. $\tilde{\mathcal{H}}$  is a tensor product of enlarged Hilbert spaces of individual sites  $\tilde{\mathcal{H}}_{i}$,  which are spanned by states $\{ |s_i, a_i\rangle, s_i = 1\dots d, a_i = 1\dots d \}$ with an index $a_i$ numbering the ancillary degrees of freedom. To obtain $\rho(T)$ from $|\psi(T)\rangle$ one needs to trace out the  ancillary degrees of freedom 
\begin{equation}
\rho(T) = \textrm{Tr}_{a} |\Psi(T)\rangle \langle \Psi(T) |. 
\label{rho}
\end{equation}  
For  $T=\infty$ we have 
\begin{equation}
 |\Psi(T=\infty)\rangle \propto  \bigotimes_i \big( \sum_{s_i=1\dots d} |s_i s_i \rangle \big), 
\label{puri0}
\end{equation}
and for  finite  $T$ $|\Psi(T)\rangle$ is obtained by an action of  $e^{-H/(2T)}$ on the physical degrees of freedom of  ${|\Psi(T=\infty)\rangle}$ 
\begin{equation}
 |\Psi(T)\rangle \propto e^{-H/(2T)}  |\Psi(T=\infty)\rangle. 
\label{puri}
\end{equation} 

\section{iPEPS representation of thermal states}
\label{sec:iPEPS}

\begin{figure}[]
\begin{center}
  \includegraphics[width= \columnwidth]{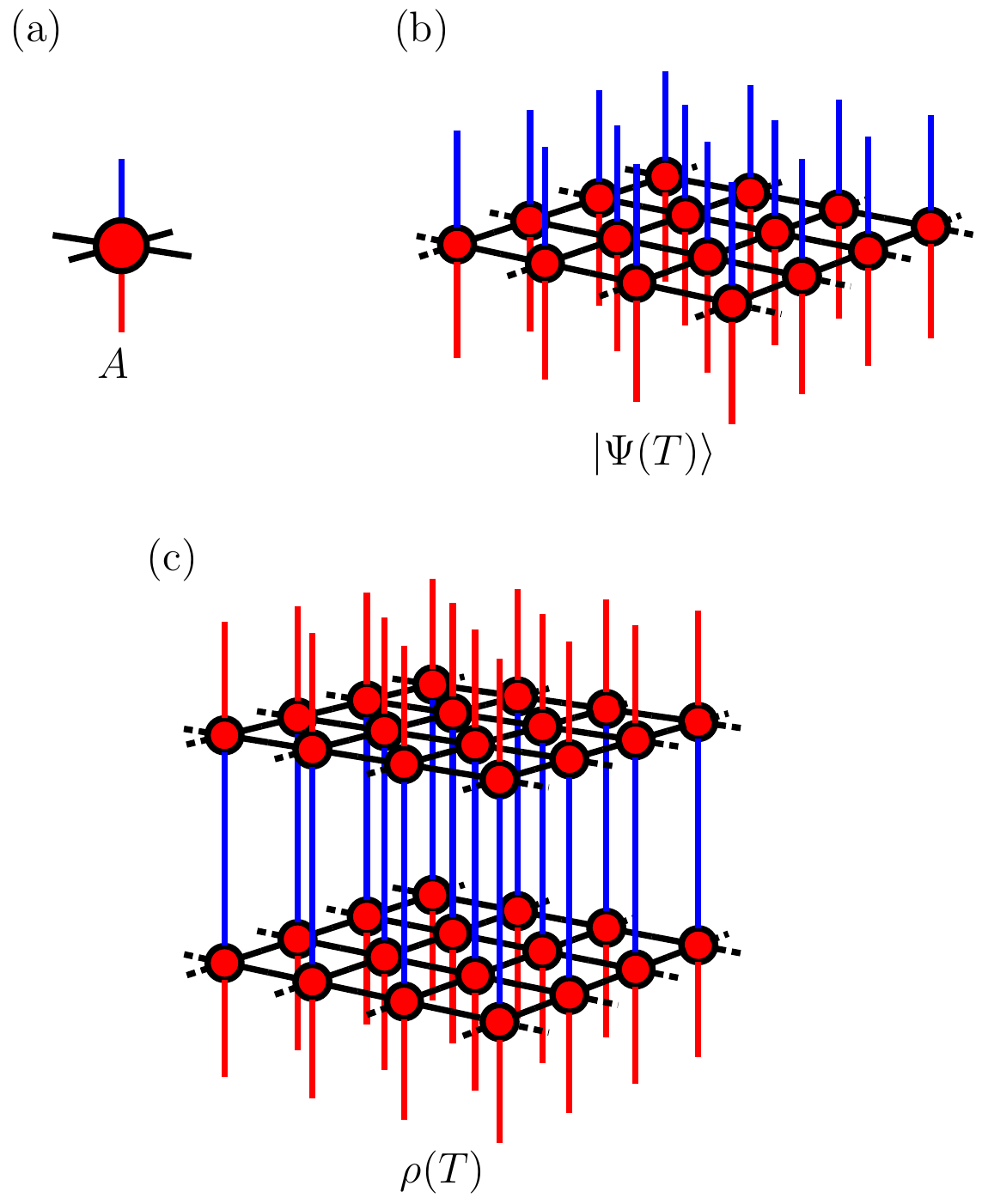}
\end{center}
  \caption{In (a) a rank 6 PEPS tensor $A$ for an iPEPS representation of a thermal state's purification $|\Psi(T)\rangle$. Each leg corresponds to an index of the tensor. The black indices are  virtual indices with  bond dimension $D$.  The red index numbers physical degrees of freedom and the blue index numbers  ancillary  degrees of freedom. In (b) the iPEPS representation of $|\Psi(T)\rangle$ is built of copies of $A$ whose  virtual indices were  contracted. Here we show the case of a translationally invariant $|\Psi(T)\rangle$. The dashed black lines indicate that only a part of the infinite tensor network is shown. In (c)  $\rho(T)$  is given by a contraction of the ancillary indices of the iPEPS  representing  $|\Psi(T)\rangle$ (the lower one) and its hermitian conjugate  $\langle\Psi(T)|$ (the upper one).   }
   \label{fig:rho}
\end{figure}

A projected entangled-pair state state (PEPS) \cite{Verstraete_PEPS_04}, also called  a tensor product state  \cite{Nishino_2DvarTN_01,Nishino_2DvarTN_04}, is a 2D  tensor network representing a state obeying the area law of  entanglement.  In the simplest case, a PEPS represents a pure state and  is built from a network of rank 5 tensors on a square lattice, with one tensor per lattice site. Each tensor has a physical index representing the local Hilbert space of a site. The other four indices of a tensor, called the virtual indices,  are contracted  with the virtual indices of the neighboring tensors. Their dimension is called the bond dimension $D$ which controls the accuracy of the ansatz. With growing $D$  states with stronger entanglement can be represented by the PEPS. 
 An infinite projected entangled pair  state (iPEPS)  is a PEPS representing a state on an infinite lattice. To obtain an iPEPS we introduce a unit cell of tensors which is periodically repeated on the lattice, i.e the iPEPS is translational invariant by shifts of the unit cell size.  With each site in the unit cell we associate a different PEPS tensor. In this work we use a unit cell with two tensors $A$ and $B$ arranged in a checkerboard pattern (all states studied in this work are compatible with this unit cell). 

As $|\Psi(T)\rangle$ is a pure state in the extended Hilbert space, it can be represented by an iPEPS with the tensors having an additional index for the ancillary degrees of freedom, see Fig.~\ref{fig:rho}(a,b). Then $\rho(T)$ (\ref{rho})  can be obtained by a contraction of the iPEPS representation of $|\Psi(T)\rangle$ and its hermitian conjugate, see  Fig.~\ref{fig:rho}(c).

The state at infinite temperature,  $|\Psi(T=\infty)\rangle$, can be represented exactly by an iPEPS with the bond dimension $D=1$, i.e. a product state (\ref{puri0}).  For finite $T$ $|\Psi(T)\rangle$ is 
obtained by an action of an operator  $e^{-H/(2T)}$ on the physical degrees of freedom of  $|\Psi(T=\infty)\rangle$ (\ref{puri}). As $e^{-H/(2T)}$ in general does not have a numerically tractable exact tensor network representation, we use its Suzuki-Trotter decomposition 
 \cite{Trotter_59,Suzuki_66,Suzuki_76} and  VTNR  \cite{Czarnik_VTNR_15, Czarnik_fVTNR_16} to find an  iPEPS approximating  $|\Psi(T)\rangle$ for a given bond dimension $D$.

\section{VTNR }
\label{sec:VTNR}

\begin{figure}[]
\begin{center}
  \includegraphics[width= \columnwidth]{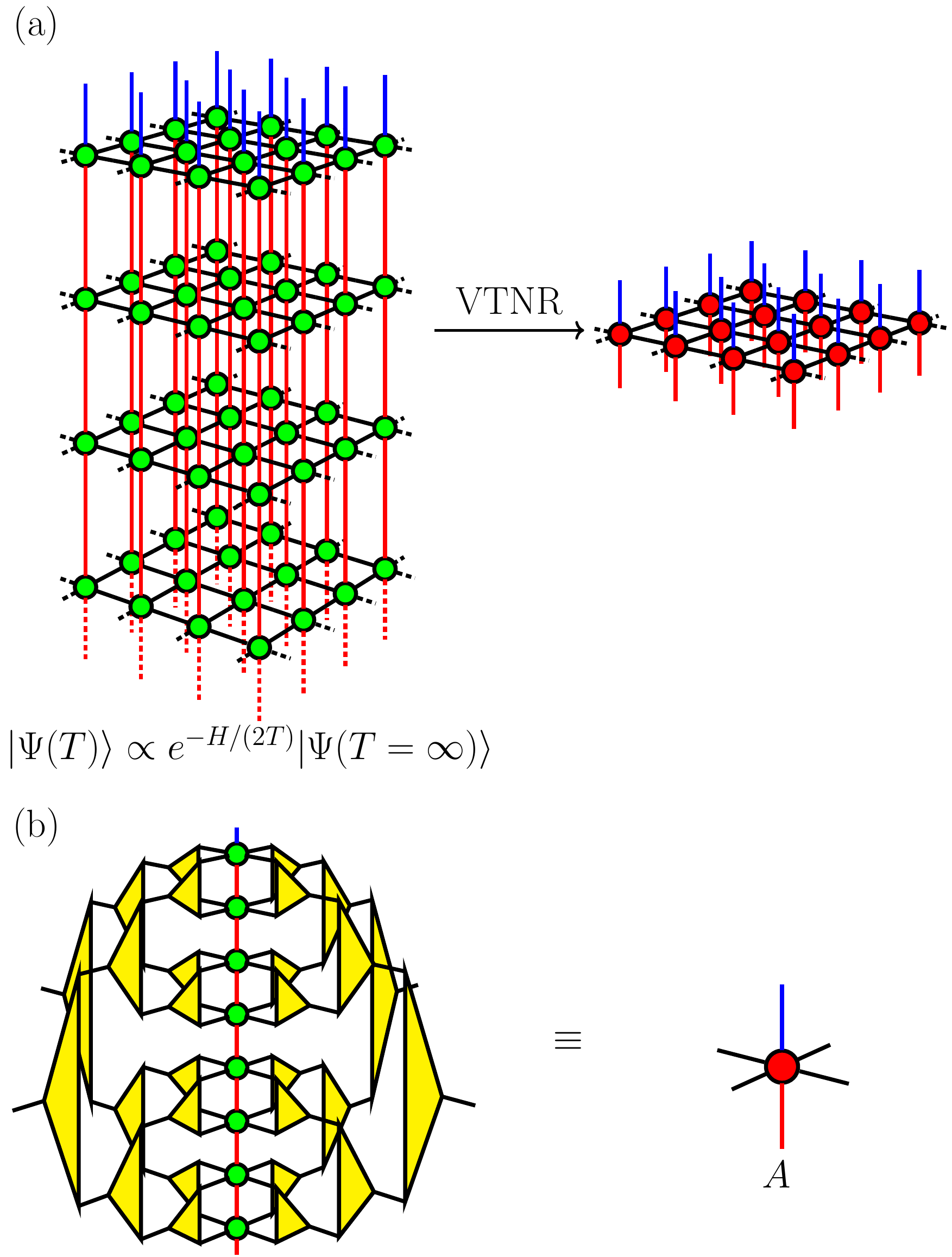}
\end{center}
\caption{ In (a), at the left a 3D tensor network approximating $|\Psi(T) \rangle \propto e^{-H/(2T)}  |\Psi(T=\infty)\rangle$ obtained by a Suzuki-Trotter decomposition (\ref{ST}) is shown.  The top layer of the tensor network is an iPEPS with $D=1$  representing $ |\Psi(T=\infty)\rangle$~(\ref{puri0}). The lower layers  are   iPEPO representations of the exponentials of classical Hamiltonians appearing in the Suzuki-Trotter decomposition (\ref{ST}) of the $e^{-H/(2T)}$ operator, which    acts on the physical indices (the red ones) of $ |\Psi(T=\infty)\rangle$.  
VTNR approximates  the 3D network by an iPEPS  with the numerically tractable $D$ shown  at the right.  Here we show the case of a translationally invariant  $|\Psi(T) \rangle$.    In (b) the PEPS tensor $A$    is obtained by VTNR from tensors of the 3D tensor network. Tree tensor networks of isometries (the yellow ones) are applied to virtual indices (the black ones) of the 3D tensor network giving $A$. Here we show the case of the 3D network build from eight layers. }

   \label{fig:VTNR}
\end{figure}

We treat an operator  $e^{-H/(2T)}$ as an imaginary time evolution operator $e^{-\tau H}$ with the imaginary time  $\tau = 1/(2T)$. We decompose $H$ into a sum of classical Hamiltonians $H_{cl}^{j}$, i.e Hamiltonians which are sums of commuting terms,
\begin{displaymath}
H = \sum_{j=1,\dots,m} H_{cl}^{j}.
\end{displaymath}
We use a second order Suzuki-Trotter decomposition \cite{Trotter_59,Suzuki_66,Suzuki_76} to approximate $e^{-\tau H}$ 
\begin{eqnarray}
e^{-\tau H} = 
&\big( e^{-\delta \tau/2 H_{cl}^{1}} 
 \dots e^{-\delta \tau/2 H_{cl}^{m-1}}  e^{-\delta \tau H_{cl}^{m}}  \nonumber\\
& e^{-\delta \tau/2 H_{cl}^{m-1}} \dots e^{-\delta \tau/2 H_{cl}^{1}} \big)^{\tau/\delta\tau} + O(\delta \tau^2). 
\label{ST}
\end{eqnarray}
The accuracy of the decomposition (\ref{ST}) is controlled by the size of the small time step $\delta \tau$.
An exact iPEPO representation of an exponential of  a classical Hamiltonian with a finite range of interaction  can be found analytically and has $D=d^2$ at most (see Ref.~\onlinecite{Wolf_Tarealaw_08} and simple examples in Refs.~\onlinecite{Czarnik_VTNR_15, Czarnik_fVTNR_16}).  Using the Suzuki-Trotter decomposition (\ref{ST}) and iPEPO representations of the exponentials of $H^{j}_{cl}$, we approximate $|\Psi(T)\rangle$ (\ref{puri}) by the 3D tensor network shown in Fig.~\ref{fig:VTNR}(a). 

 We use VTNR to approximate the 3D tensor network  by an iPEPS  with a numerically tractable bond dimension $D$, which yields an approximate representation of  $|\Psi(T) \rangle$. In VTNR the iPEPS tensors are obtained by  acting with tree tensor networks consisting of  isometries~\cite{Tagliacozzo_TTN_09} on the virtual indices of the 3D tensor network, as shown in Fig.~\ref{fig:VTNR}(b). The isometries are found by a variational update to minimally distort the partition function $Z(T)$ as described in detail in Refs.~\onlinecite{Czarnik_VTNR_15, Czarnik_fVTNR_16}. The accuracy of the final iPEPS is controlled systematically by the bond dimension of the isometries, which here equals the bond dimension of the final iPEPS.

\section{Corner Transfer Matrix Renormalization Group (CTMRG) }
\label{sec:CTM}

\begin{figure}[htb]
\begin{center}
  \includegraphics[width= \columnwidth]{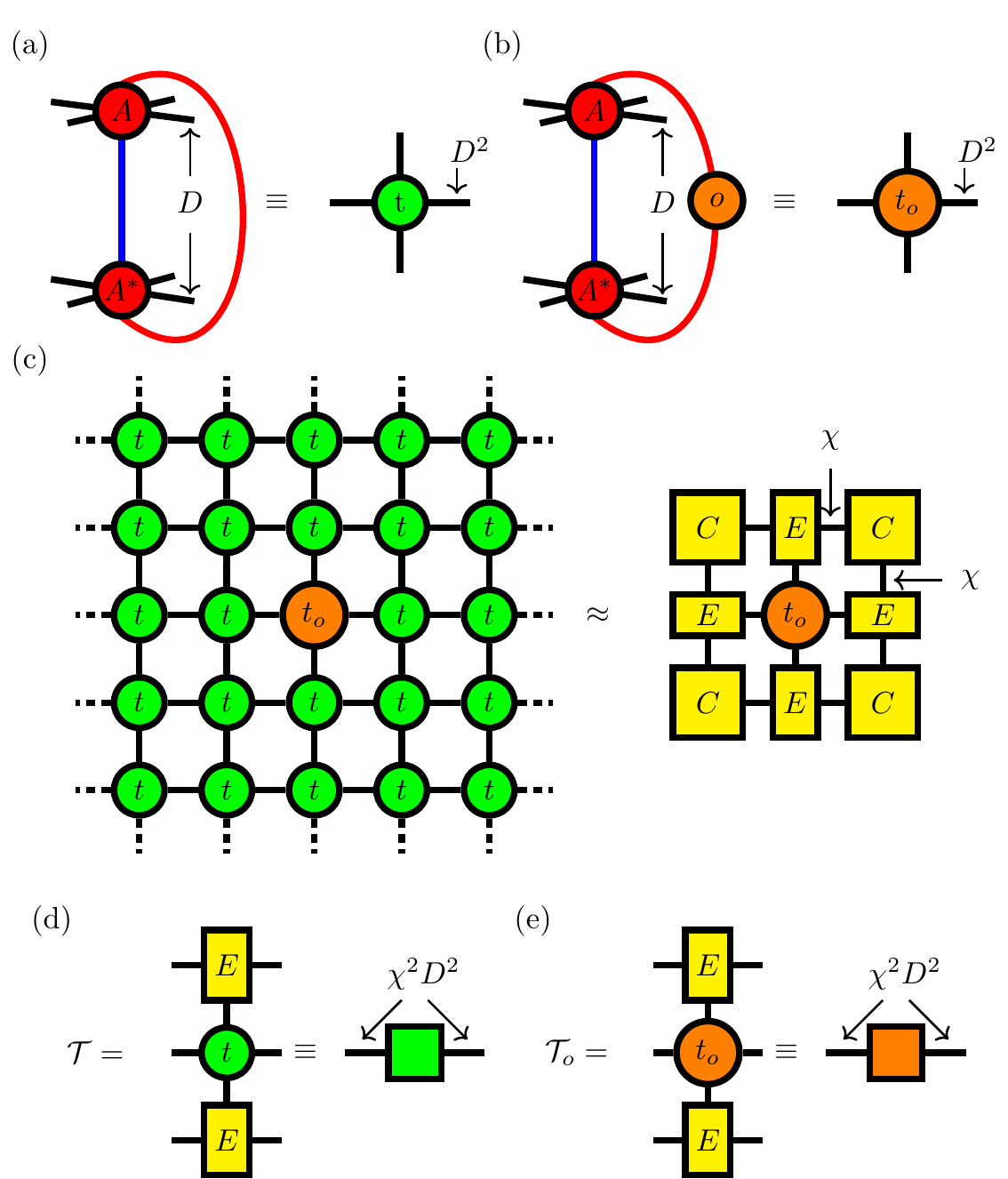}
\end{center}
  \caption{In (a) physical (blue)  and ancillary (red) indices of a PEPS tensor $A$ and its complex conjugate $A^{*}$   are contracted to create a tensor $t$. Furthermore each pair of corresponding virtual indices of the contracted tensors, which have  bond dimension $D$,  is treated as a  virtual index of the tensor $t$ with bond  dimension $D^2$. In (b) a tensor $t_o$ is created analogously to $t$ for a site at which an operator $o$ is acting.  In (c) the diagram used to compute an expectation value of $o$   by the Corner Transfer Matrix Renormalization Group (CTMRG) is shown. CTMRG approximates the infinite tensor network shown  on the left  by a contraction of a finite tensor network of  tensors $C$, $E$ and $t_o$. The accuracy of the approximate contraction is systematically controlled by the  bond dimension $\chi$ of the environment tensors $C$ and $E$. In (d) the CTMRG transfer matrix $\mathcal{T}$, which is used to compute a correlation length $\xi$ of $\rho(T)$ is shown. By grouping  the three left and the three right indices of the transfer matrix into  single indices we obtain a matrix which we use to compute $\xi$ (see Sec.~\ref{sec:extrap}).  In (e) we define, analogously to (d), an operator $o$ transfer matrix $\mathcal{T}_o$ which is used in the $\xi$ extrapolation procedure, see Sec.~\ref{sec:extrap}.   In (c,d,e) for simplicity   we show CTMRG for the  translationally  invariant $|\Psi(T)\rangle$. Furthermore we assume a translational invariant state with tensor  $A$ being rotational and mirror symmetric.}
   \label{fig:CTM}
\end{figure}

To compute expectation values of observables and the correlation length $\xi$ we use CTMRG \cite{Baxter_CTM_78,Nishino_CTMRG_96,Orus_CTM_09,Corboz_CTM_14}. CTMRG approximates contractions of an infinite  number of copies of the PEPS tensors by  contractions of a finite number of environment tensors $C, E$,  where the accuracy is systematically controlled by the  bond dimension $\chi$ of the environment tensors. An example of such an approximation is shown in Fig.~\ref{fig:CTM}(a,b,c). Details of the algorithm can be found in Refs.~\onlinecite{Corboz_CTM_14, Czarnik_fVTNR_16}. 

\section{$\xi$ extrapolation}
\label{sec:extrap}

The correlation length $\xi$ converges very slowly with increasing $\chi$, unlike  local observables, e.g.  the energy or the magnetization \cite{Rams_xiextrap_18, Corboz_FCLS_18}. 
Therefore, to determine the correlation length $\xi$ of  $|\psi(T)\rangle$  we use an extrapolation procedure from Ref.~\onlinecite{Rams_xiextrap_18} which  we summarize briefly below in the simplest case of a translationally invariant PEPS. The procedure uses the eigenvalues  of the CTMRG  transfer matrix $\mathcal{T}$, see  Fig.~\ref{fig:CTM}(d). 
  
 To  set up the extrapolation we define $\epsilon_j$ for each eigenvalue $\lambda_j$  of  $\mathcal{T}$,
\begin{equation}
\lambda_j/\lambda_0 = e^{-(\epsilon_j + i \phi_j)}, 
\end{equation}  
where $j$ numbers the  eigenvalues of  $\mathcal{T}$ ordered by the absolute value
$|\lambda_0| \ge |\lambda_1| \ge   |\lambda_2| \dots$ and $\phi_j \in (-\pi,\pi]$ determines the phase of $\lambda_j$.   A connected correlation function of an one-site operator $o$  at a distance $R$ is then expressed as 
\begin{equation}
C_{oo}(R) = \langle o_0 o_R \rangle - \langle o_0 \rangle \langle o_R \rangle = \sum_{j>0} 
f_j^{oo}  e^{-(\epsilon_j + i \phi_j) R}.
\end{equation}
Here  the form factors $f_j^{oo}$  are defined  as
\begin{equation}
f_j^{oo} = (\phi_0|\mathcal{T}_o|\phi_j) (\phi_j| \mathcal{T}_o | \phi_0),
\end{equation} 
$\mathcal{T}_o$ is a transfer matrix of the operator $o$ shown in Fig.~\ref{fig:CTM}(e), and 
$| \phi_j)$, $( \phi_j|$ are left and right eigenvectors of $\mathcal{T}$ normalized as $(\phi_i|\phi_j)=\delta_{ij}$. Therefore the correlation length obtained from $\mathcal{T}$ equals 
\begin{displaymath} 
\xi_{\mathcal{T}} = 1/\epsilon_1.
\end{displaymath}

\begin{figure}[h]
\begin{center}
  \includegraphics[width= \columnwidth]{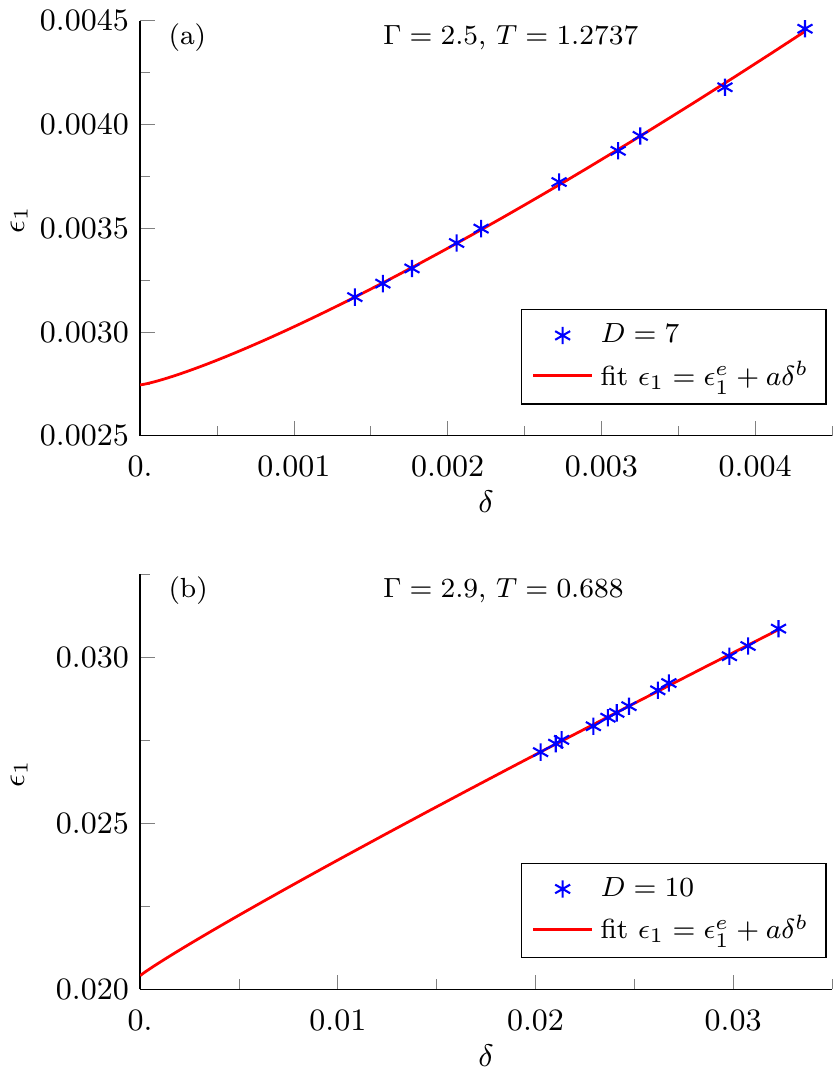}
\end{center}
  \caption{ Examples of the $\xi$ extrapolation for the Quantum Ising model in the vicinity of $T_c$. In (a),  the case of  $\Gamma = 2.5$ and $T= 1.2737$ simulated  with $D=7$ and $56 \ge \chi \ge 168$.  The fit gives the extrapolated inverse $\xi$  $\epsilon_1^{e}=0.00275(10)$ and the exponent $b =1.23(14)$.
In (b), the extrapolation  for $\Gamma = 2.9$ and  $T=0.6088$ simulated with $D=10$ and $85 \ge \chi \ge 220$. Here the fit  gives  $\epsilon_1^{e}=0.020(3)$ and  $b =0.94(35)$.}
   \label{fig:extrap}
\end{figure}

 As the spectrum of  $\mathcal{T}$ is continuous  in the limit of $\chi\to \infty$ the extrapolation uses its deviation from continuity $\delta$ as a  measure of finite $\chi$ effects. To extrapolate we use   eigenvalues of $\mathcal{T}$ contributing to a connected correlation function of the phase transition's order parameter $C^{mm}(R)$, i. e. the eigenvalues with non-zero form factors of the order parameter $f_j^{mm}$.  We denote $\epsilon$'s of such eigenvalues by $\{\epsilon^{mm}_k, k = 1, 2, \dots\}$ with $\epsilon_1^{mm} \le \epsilon_2^{mm} \le \dots$.
We remark that in the case of a second order phase transition  the diverging $\xi$ is associated with the symmetry  breaking, so we expect that the leading eigenvalue of the transfer matrix determines the asymptotics  of the order parameter correlation function, i.e.  $\epsilon_1^{mm}$ = $\epsilon_1$.  We observe that this is indeed the case for the transitions investigated below. We  note that $\lambda_0$ does not contribute to  $C^{mm}(R)$ by definition.
 We define  $\delta$ as a distance in between two dominant $\epsilon^{mm}_k$, i. e.
\begin{equation}
\delta = \epsilon_2^{mm} - \epsilon_1^{mm}.
\end{equation}    
We note that this choice of $\delta$ was proposed and benchmarked in Ref.~\onlinecite{Rams_xiextrap_18}.    
Using the spectra from different values of $\chi$, we extrapolate $\epsilon^{mm}_1$ as a function of $\delta$  by fitting 
\begin{equation}
\epsilon^{mm}_1 = \epsilon_1^{e} + a \delta^b,
\end{equation}   
 where $\epsilon_1^{e}$ is an extrapolated value of $\epsilon^{mm}_1$ and $a, b$ are parameters of the fit. The extrapolation gives us  
\begin{equation}
\xi = 1/\epsilon_1^{e}.
\end{equation}
 In Fig. \ref{fig:extrap} we present examples of the extrapolation for the Quantum Ising model with  parameters which are  investigated later in Sec. \ref{sec:QIs}. 
  
\section{Finite correlation length scaling}
\label{sec:scaling}
\label{sec:FCLS}

Finite Correlation Length Scaling (FCLS) was introduced for simulations of 1D  quantum critical phenomena \cite{Tagliacozzo_FES_08,Pollmann_FES_09,Pirvu_FES_12} with infinite MPS.  In this case a finite iMPS bond dimension $D$ introduces a finite correlation length $\xi_D$ at a critical point. $\xi_D$ acts as a cutoff on the diverging correlation length, similarly as a finite system size. It was shown that a scaling analysis in $\xi_D$ can be done in a similar way as in conventional finite size scaling, by replacing the system size $L$ by $\xi_D$ \cite{Tagliacozzo_FES_08,Pollmann_FES_09,Pirvu_FES_12} in a scaling ansatz, and then make use of this ansatz to obtain the location of the critical point and  the values of  universal critical exponents.  As the finite $D$ introduces also a finite entanglement entropy at the critical point~\cite{Vidal_1Dcrit_03} FCLS for 1D critical phenomena  was originally called  Finite Entanglement Scaling  \cite{Tagliacozzo_FES_08,Pollmann_FES_09}.
The FCLS was recently   applied to iPEPS  simulations of Lorentz-invariant quantum critical points \cite{Corboz_FCLS_18,Rader_FCLS_18} for which it was found that a finite $D$ introduces finite $\xi_D$ at the critical point \cite{Corboz_FCLS_18}.      

Here we consider a second order finite temperature phase transition for a quantum Hamiltonian. We use VTNR  to find  finite $D$ iPEPS approximating  purifications of thermal states in the vicinity of the critical temperature $T_c$. We observe that the VTNR optimization introduces a finite correlation length $\xi_D$ at $T=T_c$ (or equivalently at  $t = (T-T_c)/T_c = 0$) for all $D$'s reached in this work
\begin{equation}
\xi(t=0,D) = \xi_D. 
\end{equation} 
This observation motivates us to consider an application of FCLS to obtain $T_c$ and the  critical exponents from the VTNR results.  

We obtain  the FCLS ansatz for observables  from the standard Finite Size Scaling ansatz by replacing the finite system size  by   $\xi_D$, e. g. for the order parameter $m(t,D)$ we use
\begin{equation}
m(t,D) = \xi_D^{-\beta/\nu}\mathscr{M}(t\xi_D^{1/\nu}),
\label{mscal}
\end{equation} 
where $\beta$, $\nu$ are the critical exponents, and $\mathscr{M}$ is a non-universal  function.

To compute observables  we  contract the  iPEPS using CTMRG. The finite $\chi$  introduces  an effective length-scale $\xi_{\chi}$ ~\cite{Nishino_scalCTM_96}, i.e finite $\chi$ values of the observables are given by an more complicated scaling ansatz depending on both $\xi_{D}$ and $\xi_{\chi}$,
 e. g. 
\begin{equation}
m(t,D,\chi) = \xi_D^{-\beta/\nu}\bar{\mathscr{M}}(t\xi_D^{1/\nu},\xi_D/\xi_{\chi}),
\label{mscalD}
\end{equation} 
where $\bar{\mathscr{M}}$ is another non-universal  function. To avoid working with the more complicated ansatz we work in the limit of $\chi \to \infty$ as proposed in Refs.~\onlinecite{Corboz_FCLS_18,Rader_FCLS_18}. We  observe
that there is no need to extrapolate  the order parameter in  $\chi$ as it converges quickly.  On the other hand, to obtain a good estimate of $\xi(t,D) \equiv \xi(t,D,\chi\to\infty)$ we use the $\xi$ extrapolation described in Sec.~\ref{sec:extrap}.   

\section{$T_c$ estimation }
\label{sec:Tc}

Finite Size Scaling usually makes use of the Binder cumulant to locate the critical point without prior knowledge of its critical exponents, but in the case of iPEPS  computation of the Binder cumulant is challenging because the 4th-order moment of the order parameter would need to be computed. Instead we apply  the $m'/m$ collapse introduced in  \cite{Corboz_FCLS_18},  which makes use of the derivative of the order parameter $m'(t,D) = dm(t,D)/dt$, to find $T_c$. A FCLS  scaling ansatz for  $m'$ is
\begin{equation}
m'(t,D) = \xi_D^{-(\beta-1)/\nu}\mathscr{M'}(t\xi_D^{1/\nu}),
\label{mderivscal}
\end{equation}
where $\mathscr{M'}$ is a non-universal  function \cite{Corboz_FCLS_18}. Using (\ref{mscal}) and (\ref{mderivscal}) we obtain 
\begin{equation}
t\frac{m'(t,D)}{m(t,D)} = \bar{\mathscr{P}}(t\xi_D^{1/\nu}),
\label{i1}
\end{equation}
\begin{equation}
\xi_D^{1/\nu} \propto \frac{m'(t=0,D)}{m(t=0,D)}, 
\label{i2}
\end{equation}
where $\bar{\mathscr{P}}$ is another non-universal  function. Eqs. (\ref{i1},\ref{i2}) give us the $m'/m$ collapse
\begin{equation}
t\frac{m'(t,D)}{m(t,D)} = \mathscr{P}\Big(t\frac{m'(t=0,D)}{m(t=0,D)}\Big), 
\label{collmderrm}
\end{equation}
We estimate $T_c$ by  plotting $y= t\frac{m'(t,D)}{m(t,D)}$ versus $x= t\frac{m'(t=0,D)}{m(t=0,D)}$ for different choices of  $T_c$ and finding  the one for which $y(x)$ data points obtained with different $D$  collapse best onto a single curve.

\section{quantum Ising model - benchmark results} 
\label{sec:QIs}

The quantum Ising model is given by the Hamiltonian: 
\begin{equation}
H = -\sum_{<i,j>} \sigma_z^i \sigma_z^j + \Gamma \sum_i \sigma_x^i,
\label{Is}
\end{equation}
where $\sigma_z$, $\sigma_x$ are Pauli matrices. 
For $\Gamma = 0$ the model   reduces to the classical Ising model with $T_c=2/\textrm{ln}(1+\sqrt{2})\approx 2.269$  and for $\Gamma_c= 3.04438(2)$ \cite{Deng_QIshc_02}  it has a quantum critical point. 
 For $0 \le \Gamma < \Gamma_c$ it exhibits a low temperature ferromagnetic phase with the order parameter $m = \langle \sigma_z \rangle$, which is separated from the paramagnetic phase by a line of finite temperature  second order phase transitions belonging to a 2D classical Ising  universality class. 

As $\Gamma$ is  approaching  $\Gamma_c$,  quantum fluctuations are becoming stronger and  $T_c$ gets  suppressed  w.r.t. the classical case of $\Gamma=0$. Therefore we expect that with increasing $\Gamma$ the accurate simulation of the finite temperature  transition  using tensor networks is becoming more challenging since a larger $D$ is necessary to correctly capture the stronger quantum fluctuations. To examine this more closely we investigate  in the following  $\Gamma = 2.5$ as well as a point close to the quantum critical point $\Gamma  =2.9$, for which  Quantum Monte Carlo (QMC) \cite{Wessel_TQMC_16} gives $T_c=1.2737(6)$ and $T_c=0.6085(8)$, respectively, corresponding to a reduction in $T_c$  with respect to $\Gamma=0$ by a factor of 1.8 and 3.7, respectively. 

\subsection{$\Gamma = 2.5$}

\label{sec:gamma2p5}

\begin{figure} [htb]
\begin{center}
  \includegraphics[width= \columnwidth]{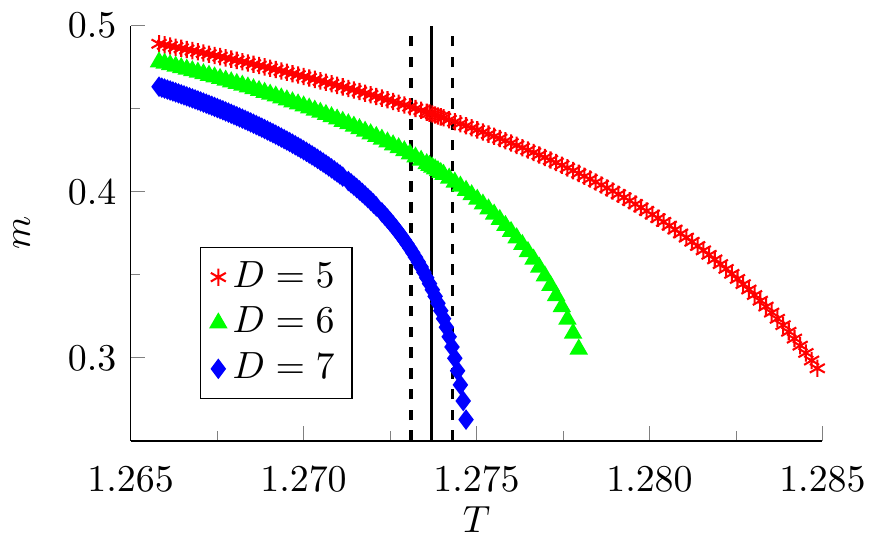}
\end{center}
  \caption{The order parameter $m$ in the vicinity of the critical temperature  for the 2D quantum Ising model with $\Gamma=2.5$ and $D=5-7$. The critical  behavior is smoothed out by finite $D$ effects. The black lines indicate the QMC estimate of the critical temperature, $T_c=1.2737(6)$~\cite{Wessel_TQMC_16}.}
   \label{fig:ord}
\end{figure}

We first consider a case well away from the quantum critical point, $\Gamma=2.5$, to provide a proof of principle of the applicability of FCLS to finite temperature VTNR simulations.  In Fig.~\ref{fig:ord} we present data for the order parameter as a function of temperature in the vicinity of the critical temperature, for bond dimensions $D=5-7$. As expected, we do not obtain a sharp phase transition but we see that the order parameter is systematically reduced with increasing $D$ similarly to the case of finite size effects. 

\begin{figure} [htb]
\begin{center}
  \includegraphics[width= \columnwidth]{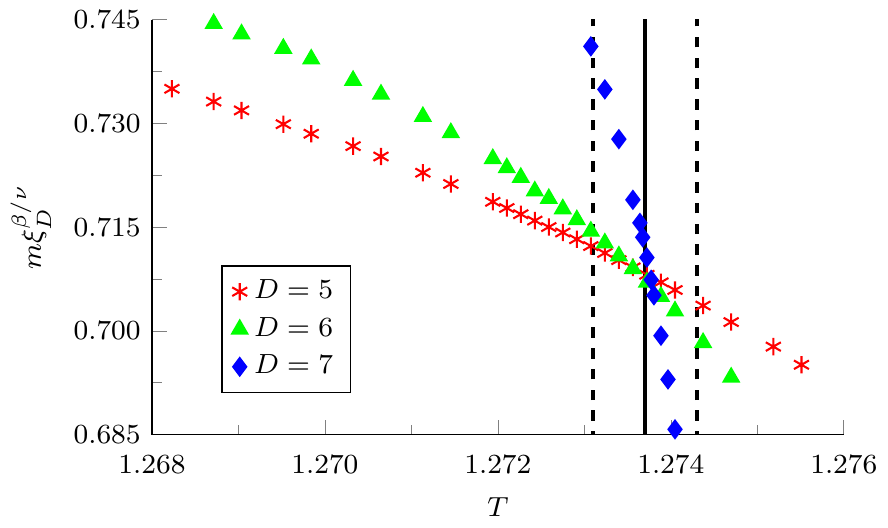}
\end{center}
  \caption{$T_c$ estimation for $\Gamma=2.5$ by intersecting  $m(T,D)\xi_D^{\beta/\nu}$  for different values of $D$  (\ref{mxid}). Here we assume the 2D classical Ising universality class with $\beta = 1/8$ and $\nu=1$. The $D=5-7$ curves intersect, as predicted by FCLS, at $T_c=1.2737(2)$ in agreement with the  QMC estimate  $T_c=1.2737(6)$~\cite{Wessel_TQMC_16}, which  is indicated by the black lines. Details of the $T_c$ estimation can be found in the text. }    
   \label{fig:mxid_scalvar}
\end{figure}

We first attempt to estimate $T_c$ by using the known critical exponents of the 2D classical Ising universality class, i.e.  $\beta=1/8$ and  $\nu = 1$. 
To do that we plot $m\xi_D^{\beta/\nu}$  which, according to FCLS, should not depend on $D$ at~$T=T_c$:
\begin{eqnarray}
m(t,D)\xi_D^{\beta/\nu} =\mathscr{M}(t\xi_D^{1/\nu}),\\
  m(t=0,D)\xi_D^{\beta/\nu} = a,
\label{mxid}
\end{eqnarray}
where $a$ does not depend on $D$. Indeed the  $m(t, D) \xi_D^{\beta/\nu}$ curves for different $D$'s cross as predicted by FCLS, see Fig.~\ref{fig:mxid_scalvar}.  We identify $T_c$ as the temperature for which the variance of   $m(t,D)\xi_D^{\beta/\nu}$ is  smallest, obtaining $T_c=1.2737(2)$ in agreement with the  QMC estimate $T_c=1.2737(6)$~\cite{Wessel_TQMC_16}. The $T_c$ uncertainty is obtained by varying the range of $D$ and the range of $\chi$ used for $\xi_D$ estimation by  extrapolation.

\begin{figure}[htb]
\begin{center}
  \includegraphics[width= \columnwidth]{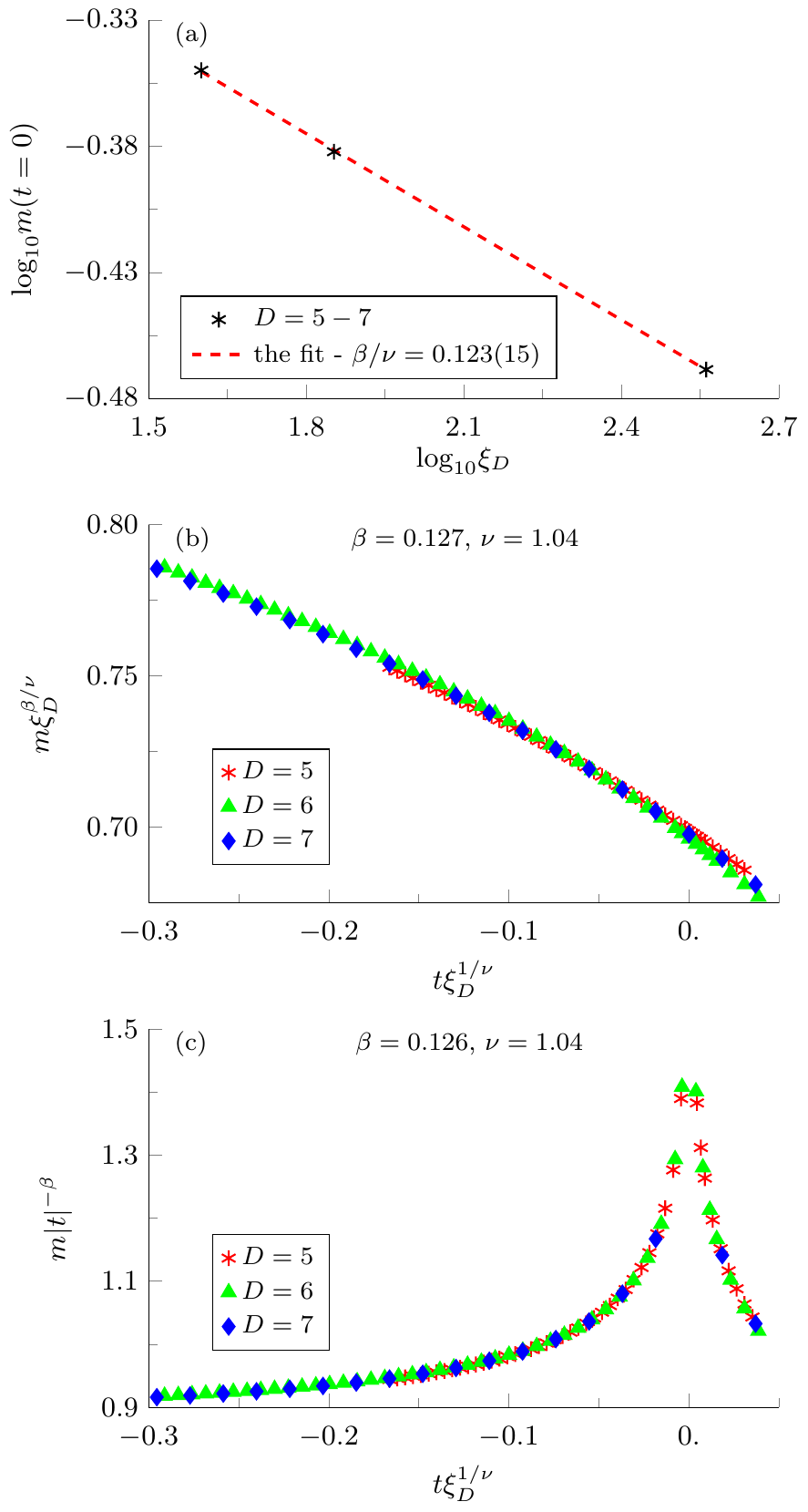}
\end{center}
  \caption{ Critical exponents  estimation for $\Gamma=2.5$, using  $T_c=1.2737(2)$ found earlier. In (a) a fit to Eq.~\eqref{Tcfit}  gives $\beta/\nu=0.123(15)$, in agreement with the exact $\beta/\nu=1/8$. In (b) and (c), by performing data  collapses using Eqs. \eqref{collmxi} and \eqref{collmt} we obtain $\beta=0.127(2),\,\nu=1.04(4)$ and   $\beta=0.126(5),\,\nu=1.04(5)$, respectively. The  $\beta$ and $\nu$ estimates agree with the exact $\beta=1/8$, $\nu = 1$.}
   \label{fig:univ}
\end{figure}

To provide further evidence of FCLS  we determine the critical exponents, using  $T_c=1.2737(2)$ found earlier. We first estimate  $\beta/\nu$ using the data obtained at $T=T_c$. From Eq.~\eqref{mxid}  we obtain 
\begin{equation}
\log m(t=0,D) = - \beta/\nu \log \xi_D + \log a.
\label{Tcfit}
\end{equation}  
A linear fit to the data on a log-log scale shown in Fig.~\ref{fig:univ}(a) yields $\beta/\nu=0.123(15)$, in agreement with the exact $\beta/\nu=1/8$. The error bar takes into account the $T_c$ uncertainty and the statistical error of the  fit.  

Next, we estimate $\beta$ and $\nu$  by performing data collapses based on data in the vicinity of $T_c$, using the scaling ansaetze: 
\begin{equation}
m(t,D)\xi_D^{\beta/\nu} = \mathscr{M}(t\xi_D^{1/\nu}),
\label{collmxi}
\end{equation}
\begin{equation}
m(t,D) t^{-\beta} = \bar{\mathscr{M}}(t\xi_D^{1/\nu}).
\label{collmt}
\end{equation}
Using the ansatz (\ref{collmxi}) we obtain $\beta = 0.127(2)$, $\nu = 1.04(4)$, in agreement with the exact universality class, see Fig.~\ref{fig:univ}(b). Ansatz (\ref{collmt}) yields  $\beta = 0.126(5), \nu = 1.04(5)$, again in agreement with the exact exponents, see Fig.~\ref{fig:univ}(c). In both cases the uncertainties are obtained by taking into account the $T_c$ uncertainty and by varying the data range.

\begin{figure} [htb]
\begin{center}
  \includegraphics[width= \columnwidth]{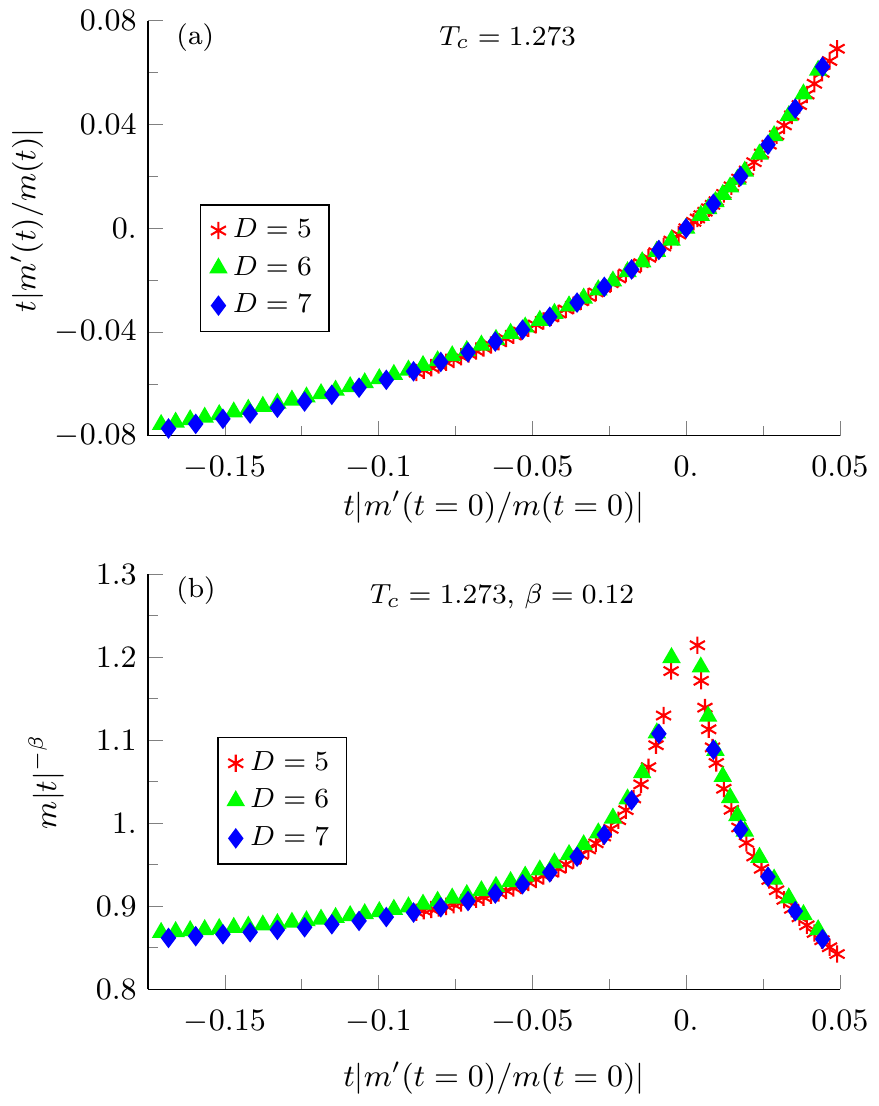}
\end{center}
  \caption{$T_c$ and $\beta$ estimation for $\Gamma=2.5$ without knowledge of the universality class nor the value of $T_c$. In (a) we use the $m'/m$ ansatz (\ref{collmderrm}) to perform a data collapse, obtaining $T_c= 1.273(1)$.  In (b)  we estimate  $\beta$  by   performing a  data collapse based on Eq.~\eqref{collmt2}, using the value of $T_c$ found in (a), which yields $\beta=0.12(1)$. The obtained $T_c$ and $\beta$ agree with the QMC estimate $T_c=1.2737(6)$~\cite{Wessel_TQMC_16} and the exact $\beta = 1/8$. }
   \label{fig:mderivm_coll}
\end{figure}

Finally, we show that we can estimate $T_c$ and $\beta$  without prior knowledge of the universality class nor the value of $T_c$.  First, we estimate $T_c$ by performing a data collapse using the $m'/m$ ansatz (\ref{collmderrm}), which yields $T_c = 1.273(1)$, see Fig.~\ref{fig:mderivm_coll}(a). The $T_c$  uncertainty is obtained by varying the data range.   We estimate  $\beta$ by performing a data collapse based on the ansatz
\begin{equation}
m(t,D) t^{-\beta} =    \bar{\mathscr{P}}\Big(t\frac{m'(t=0,D)}{m(t=0,D)}\Big),
\label{collmt2}
\end{equation}
using for  $T_c$ the value  obtained in the $m'/m$ collapse.  We obtain $\beta= 0.12(1)$, see Fig.~\ref{fig:mderivm_coll}(b). The uncertainty is obtained taking into account the $T_c$ uncertainty and varying the data range.  

We remark that while the  quality of the obtained results is  good,  we see some deviations from perfect  scaling which may be caused either by corrections to finite size scaling or limitations of VTNR in getting  optimal tensors. First we note that VTNR is not guaranteed to give the best iPEPS approximation of the thermal state's purification for a given $D$, as it does not search  directly for the best iPEPS tensor representing thermal state purification. Instead it optimizes a tree tensor network (TTN) of isometries which, applied to virtual indices of the tensor network representing  a Suzuki-Trotter decomposition of the purification, gives the iPEPS approximating the purification. While this approach makes the variational optimization of the iPEPS  efficient it is not equivalent to the most general iPEPS variational optimization procedure. Still our results demonstrate that the accuracy of the optimized tensors is high enough to extract the critical coupling and critical exponents with a good accuracy.

 Second  we note  that for $\Gamma=2.5$  CTMRG convergence is  challenging, as for $D=5-7$ we obtain  $\xi_{D=7}\sim 350$. For an iPEPS with such a large $\xi$   many iterations of the CTMRG procedure are necessary to converge $m$. Good convergence of the CTMRG environment is important for   the  variational optimization  since VTNR uses the CTMRG environmental tensors  to find the best iPEPS~\footnote{To ensure a good CTMRG convergence we require the change of $m$ per  CTMRG iteration to be smaller than $10^{-8}$}.   Here to  optimize the  iPEPS we use $\chi = 8D$. We check that using  $\chi = 6D$ and $\chi=7D$ we obtain results ($T_c$ and  $\beta$ obtained by the $m'/m$ collapse  and  collapse (\ref{collmt2}))  in agreement with the ones obtained with $\chi = 8D$.  Nevertheless we cannot fully exclude the possibility that finite   $\chi$ effects contribute to the observed small  deviations from the perfect collapse as simulations with larger $\chi$ would be computationally very expensive. 

To obtain  $\xi_D$ we contract the final iPEPS with  $8D \le \chi \le 24D$ and use the extrapolation procedure described in Sec \ref{sec:extrap}.  We remark that in the case of VTNR simulations obtaining convergence in the small Trotter time-step $\delta\tau$ is relatively easy as the computational cost of the simulations scales at most as $O(\textrm{log}(1/\delta\tau))$. For $\Gamma = 2.5$  we use a second order Trotter decomposition with $\delta\tau = \tau/2^{10} \le 0.001$, which is small enough to give results converged in $\delta\tau$.

\subsection{$\Gamma=2.9$}

\label{sec:gamma2p9}

\begin{figure} [htb!]
\begin{center}
  \includegraphics[width= \columnwidth]{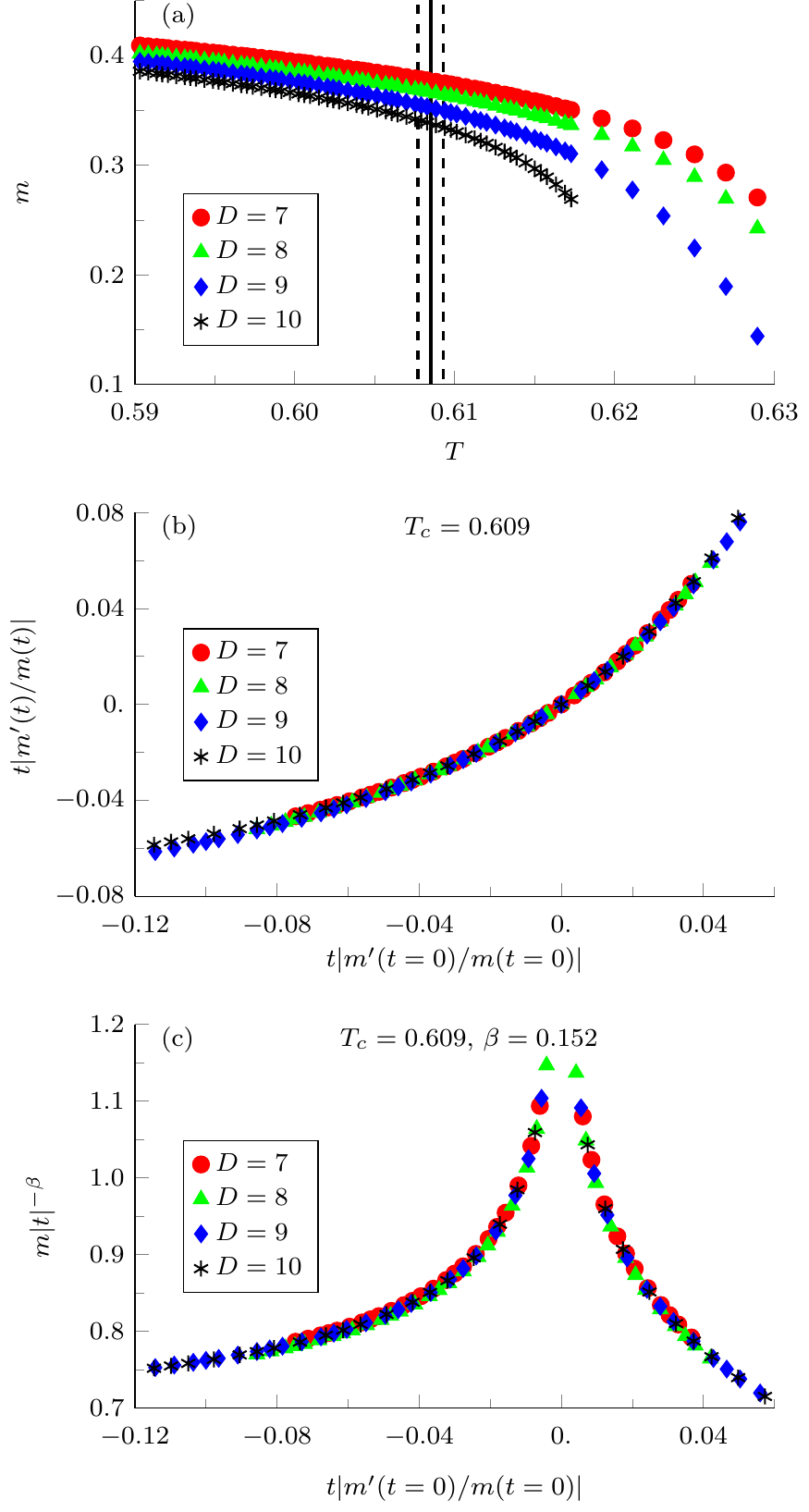}
\end{center}
  \caption{Results for the more challenging  $\Gamma=2.9$ case. In (a) the order parameter in the vicinity of $T_c$  for $D=7-10$. The black lines indicate the  QMC estimate $T_c=0.6085(8)$~\cite{Wessel_TQMC_16}. In (b) the $m'/m$  collapse (\ref{collmderrm})  gives $T_c = 0.609(4)$. In (c) $\beta$ estimation  based on a data  collapse using Eq.~\eqref{collmt2} and taking $T_c$ found in (b), gives $\beta = 0.152(8)$. The obtained $T_c$ agrees with QMC, but the obtained $\beta$ deviates from the exact one $\beta =1/8$ by about $20\%$, see main text for a discussion of this deviation.   }
   \label{fig:Gamm2p9}
\end{figure}

For the more challenging  $\Gamma=2.9$ case we analyze VTNR results for $D=7-10$, see Fig.~\ref{fig:Gamm2p9}(a). 
Using the $m'/m$  collapse we obtain $T_c = 0.609(4)$, see Fig.~\ref{fig:Gamm2p9}(b). Using this result for $T_c$ and performing a collapse with Eq.~\eqref{collmt2} we obtain  $\beta = 0.152(8)$, see Fig.~\ref{fig:Gamm2p9}(c). While the obtained  $T_c$ estimate  agrees with the QMC estimate $T_c= 0.6085(8)$~\cite{Wessel_TQMC_16}   the $\beta$ estimate deviates by about $20\%$ from the exact  $\beta =1/8$. 

Comparing the results obtained with $D=8-10$ and $D=7-10$ suggests that the $\beta$ estimate still depends significantly  on the $D$ range as we obtain $\beta = 0.145(5)$ for $D=8-10$~\footnote{Contrary to the case of $\beta$, the $T_c$ estimate for $D=8-10$,  $T_c=0.610(4)$, is similar to the $D=7-10$ estimate}.  Furthermore, we expect that the necessary $\xi_D$   to obtain the asymptotic scaling is larger  for $\Gamma= 2.9$ than  for $\Gamma= 2.5$,    as  $\Gamma= 2.9$ is closer to the quantum critical point. Despite larger $D$, the  $\xi_D$ obtained for $\Gamma= 2.9$,  although quite large ($\xi_D\sim 20 - 50$), is  smaller than $\xi_D$ for $\Gamma= 2.5$ ($\xi_D \sim 40 - 350$).  Therefore we expect that the quality of the results can  still be improved  by increasing~$D$, although it would be computationally very expensive.

The $T_c$ and $\beta$ uncertainties are estimated in the same way as in the $\Gamma=2.5$ case.   Similar values for $T_c$ and $\beta$ are obtained with VTNR  using an optimization with $\chi = 5D, 6D, 7D$.  A second order Trotter decomposition with a time step $\delta\tau = \tau/2^{10} \le 0.002$ is used.  

\section{Interacting honeycomb fermions - Benchmark results}
\label{sec:honfer}

We consider a model of interacting spinless fermions on a honeycomb lattice~\cite{Capponi_honfer_17},  given by the Hamiltonian,
\begin{equation}
H = -t \sum_{<i,j>} \Big( c_i^\dag c_j + c_j^\dag c_i \Big) + V \sum_{<i,j>} n_in_j.
\label{hon_fer}
\end{equation}
 Here $c_i(c_i^{\dag})$  is a fermionic annihilation (creation) operator at site $i$ and $n_i = c_i^{\dag}c_i$ is a fermion number operator. We set $t=1$ in the following. Furthermore, for the purpose of the benchmark we restrict ourselves to the case of  half-filling, $n = \sum_{i=1}^{N} \langle n_i\rangle/N = 1/2$, for which sign-problem free QMC results are available~\cite{Wang_honfer_14, Wang_honfer_16,Wessel_TQMC_16}.   The model has a quantum critical point at $V_c=1.356(1)$~\cite{Wang_honfer_14}. For $V>V_c$  there is a low temperature phase with  a charge density wave (CDW) order, which is separated from a disordered, 
high temperature, phase by a line of second order finite temperature  phase transitions, which  belong to the 2D classical Ising  universality class~\cite{Wang_honfer_16}. The CDW order parameter is defined as
\begin{equation}
m = \langle n_A \rangle  - \langle n_B \rangle,
\end{equation}
where  $\langle n_A \rangle$ and $\langle n_B \rangle$  are the  fermion densities on sub-lattices A and B, respectively.  In the limit of $V\to \infty$ the model becomes equivalent to the 2D classical antiferromagnetic Ising model. Here we simulate the model for $V=3$ and the more challenging case of $V=2$, which is closer to $V_c$. 

\subsection{$V=3$}
\label{sec:Vdt3}

\begin{figure} []
\begin{center}
  \includegraphics[width= \columnwidth]{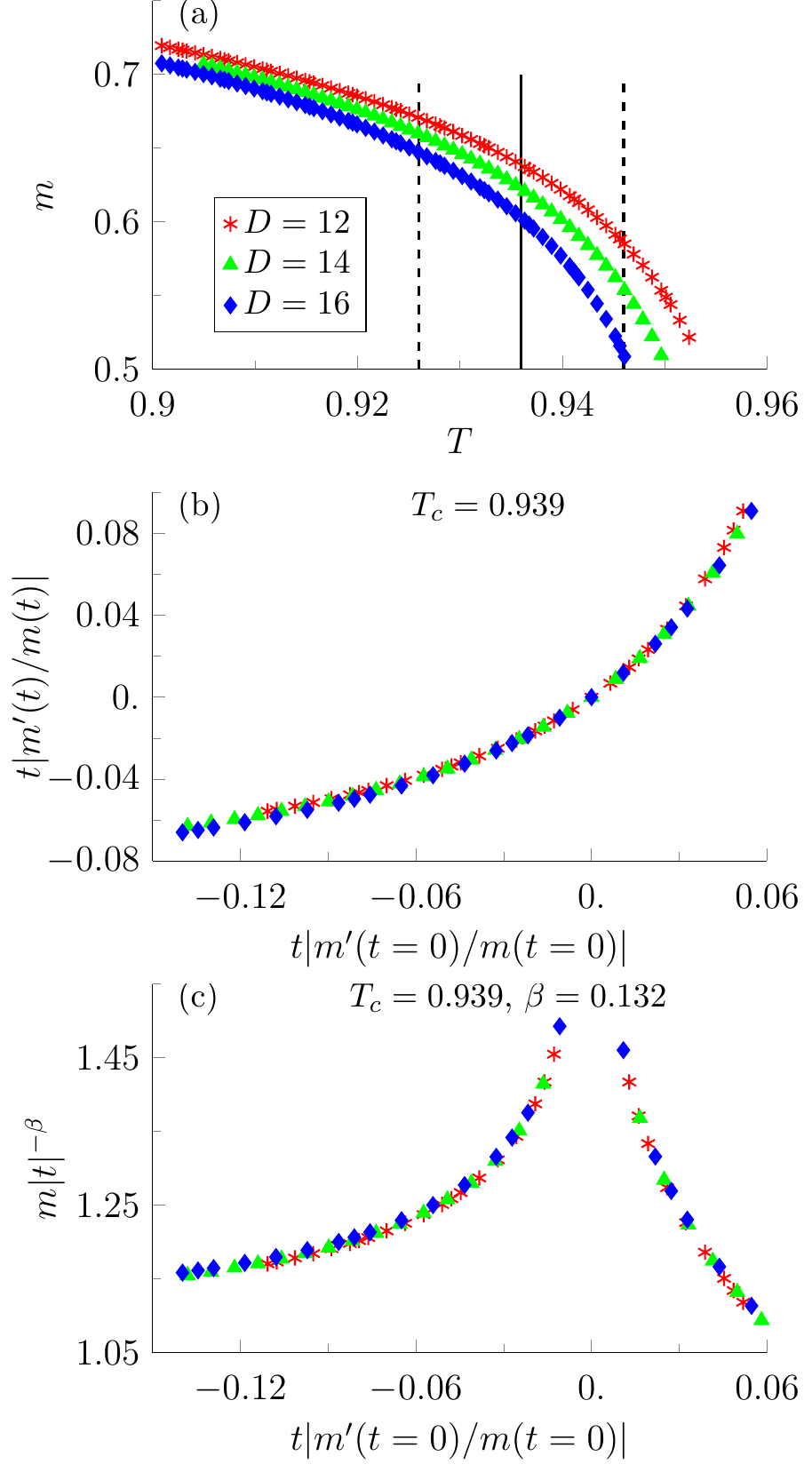} 
\end{center}
  \caption{Results for the spinless honeycomb fermion model~\eqref{hon_fer} with $V=3$. In (a) the order parameter close to $T_c$ for  $D=12,14,16$ is shown. The black lines indicate the  QMC estimate $T_c=0.936(10)$~\cite{Wessel_TQMC_16}.  In (b) $T_c$ is estimated by the  $m'/m$ collapse  giving $T_c=0.939(4)$ in agreement with QMC. In (c) a collapse using Eq.~\eqref{collmt2} yields $\beta = 0.132(8)$  using $T_c$ found in (b). The obtained $\beta$ agrees with the exact $\beta=1/8$.  }
   \label{fig:Vdt3}
\end{figure}

For $V=3$ we analyze  $D=12,14,16$ VTNR results in  the vicinity of $T_c$, see Fig.~\ref{fig:Vdt3}(a). Here QMC predicts  $T_c= 0.936(10)$~\cite{Wessel_TQMC_16}.  We determine $T_c$  using the $m'/m$ collapse obtaining $T_c= 0.939(4)$ in agreement with QMC, see Fig.~\ref{fig:Vdt3}(b). Furthermore, we obtain  $\beta=0.132(8)$ in agreement with the exact $\beta=1/8$  by performing a data collapse using Eq.~\eqref{collmt2} and by taking $T_c$ obtained from the $m'/m$ collapse, see Fig.~\ref{fig:Vdt3}(c).

The $T_c$ and $\beta$ uncertainties are obtained similarly as for the quantum Ising model.   We use $\chi =5D$  to perform the VTNR optimization obtaining results which are consistent with the ones obtained with  $\chi =3D$ and $\chi = 4D$. We use a second order Suzuki-Trotter decomposition with a time step $\delta \tau = \tau/2^{11} < 0.001$.

\subsection{ $V=2$}
\label{sec:Vdt2}

\begin{figure} []
\begin{center}
  \includegraphics[width= \columnwidth]{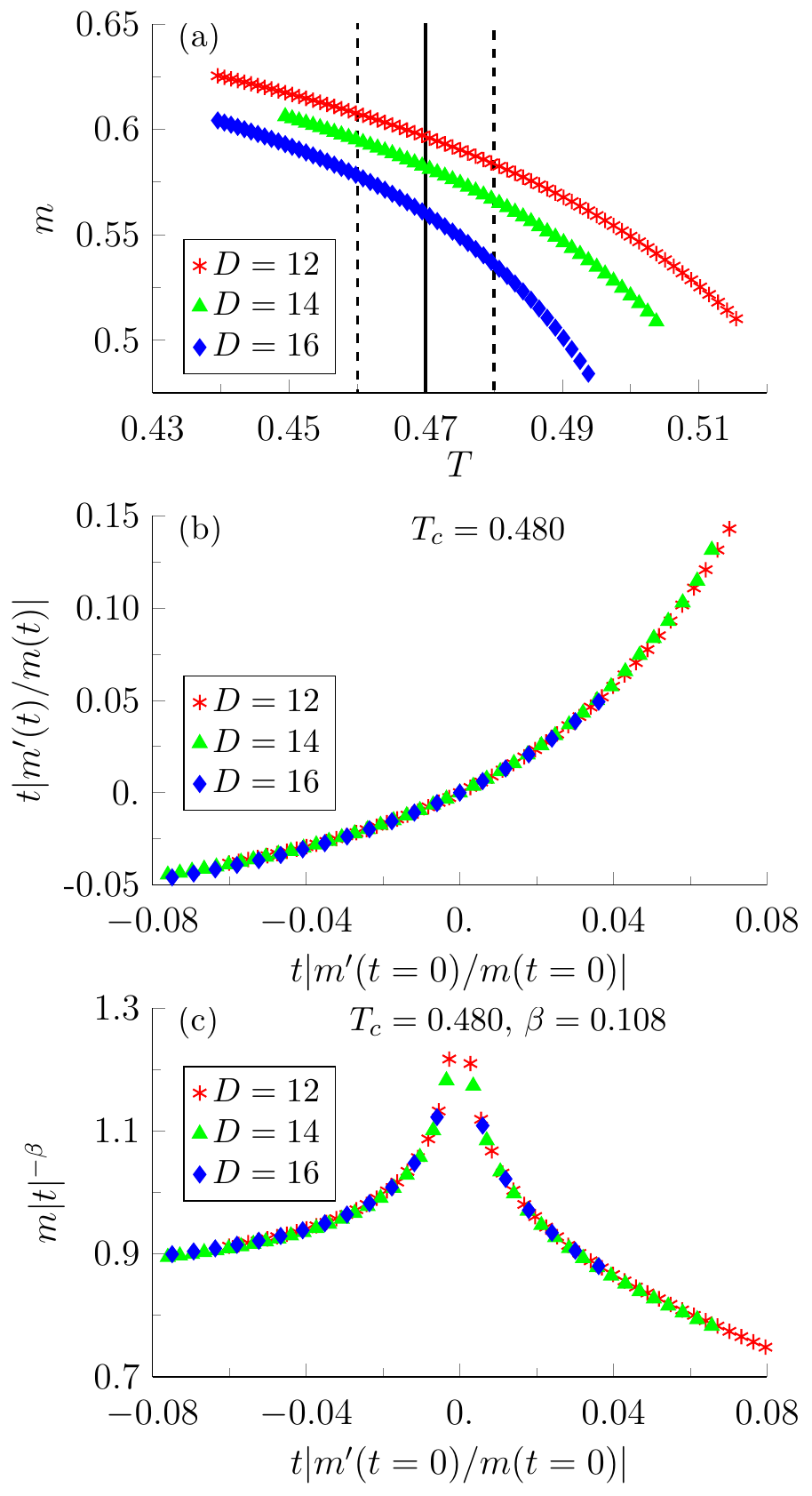} 
\end{center}
  \caption{Results for the more challenging $V=2$ case. In (a) the order parameter close to $T_c$ for $D=12,14,16$ is shown. The black lines indicate the QMC estimate $T_c=0.47(1)$~\cite{Wessel_TQMC_16}. In (b) the $m'/m$  data collapse gives $T_c = 0.480(5)$ in agreement with QMC. In (c)  a data collapse using Eq.~\ref{collmt2} yields $\beta = 0.108(4)$  using $T_c$ found in (b). The obtained   $\beta$ differs from the exact $\beta=1/8$  by  about $10\%$.  A discussion of the deviation in $\beta$  can be found in main text.}
   \label{fig:Vdt2}
\end{figure}

Next we analyze $D=12,14,16$ VTNR results  for the more challenging $V=2$ case, see Fig.~\ref{fig:Vdt2}(a). Using the $m'/m$ collapse we obtain $T_c = 0.480(5)$  in agreement with the QMC estimate $T_c=0.47(1)$~\cite{Wessel_TQMC_16},  see Fig.~\ref{fig:Vdt2}(b). Using anatz~\eqref{collmt2} and taking $T_c$ found by the $m'/m$ collapse, we obtain $\beta=0.108(4)$, see Fig.~\ref{fig:Vdt2}(c). The $\beta$ estimate deviates by about $10\%$ from the exact  $\beta = 1/8$. We see that the obtained $\xi_D$ 
for $V=2$ ($\xi_D \sim 2-4$)  is much  smaller than $\xi_D$ for $V=3$  ($\xi_{D} \sim 10-20$). Furthermore, as $V=2$ is closer to the quantum critical point than $V=3$ we expect that for $V=2$ a larger $\xi_D$ is  necessary  to be in the asymptotic scaling regime and we expect that the accuracy of $\beta$ can be improved  by increasing $D$.

The $T_c$ and $\beta$ uncertainties are obtained similarly as for the quantum Ising model. Here we use $\chi  = 5D$ to perform the VTNR optimization. Consistent results are obtained also with $\chi  = 3D$ and $\chi  = 4D$. We use  a second order Suzuki-Trotter decomposition with a time step $\delta\tau = \tau/2^8 < 0.01$.

\section{Conclusions}
\label{sec:conclusions}

In this paper  we have studied   second order finite temperature phase transitions in the 2D  quantum Ising (\ref{Is}) and interacting honeycomb fermion (\ref{hon_fer}) models using infinite projected entangled-pair states (iPEPS) to represent   thermal states. The iPEPS   were obtained by Variational Tensor Network Renormalization (VTNR). We found that  at the critical temperature $T_c$  the iPEPS correlation length $\xi_D$ is finite for the computationally accessible values of the iPEPS bond dimension $D$. Motivated by this observation we  investigated the  application of   Finite Correlation Length Scaling (FCLS)   to obtain precise values of  $T_c$  and universal  critical exponents. We found that in the vicinity of $T_c$ the  order parameter obeys well the expected behavior predicted by FCLS. 

The  two models studied in this work exhibit second order finite temperature phase transitions for the transverse
 fields $\Gamma < \Gamma_c\approx 3.04438$  and the interaction strengths $V > V_c \approx 1.356$, respectively. At $\Gamma_c$ and $V_c$  second order quantum  phase transitions occur at $T=0$.    Using FCLS  we obtained estimates of $T_c$ and the critical exponents  in agreement with the QMC results  for $\Gamma=2.5$   and $V=3$ which   are sufficiently far from the quantum critical points.  For  $\Gamma$  and $V$ approaching the quantum critical points we observed that the  magnitude of $\xi_D$   and the accuracy of the critical data  become lower for the same values of $D$.  Nevertheless we were still  able to obtain $T_c$ in agreement with the QMC results for the challenging  $\Gamma = 2.9$ and $V=2$ cases. For these couplings the values of the critical exponents  exhibit a dependence on the range of $D$ values used in the scaling analysis, suggesting that larger $D$'s are needed in order to obtain more accurate estimates of the critical exponents. 

In summary, our results further demonstrate the usefulness of tensor network simulations for quantum many-body systems at finite temperature, even for the challenging case of a finite temperature continuous phase transition, for which convergence in $D$ can typically not be reached, but which can be systematically studied using FCLS.

\acknowledgments
We thank Stephan Hesselmann and Stefan Wessel for providing us numerical values of data published in Ref.~\onlinecite{Wessel_TQMC_16} and Marek Rams for useful remarks about  the manuscript. 
This research was funded by the National Science Centre (NCN), Poland 
under project 2016/23/B/ST3/00830  and  the European Research Council (ERC) under the EU Horizon 2020 research and innovation program (grant agreement No. 677061).
\bibliographystyle{apsrev4-1}

\bibliography{FCLS.bib}

\end{document}